# On Curie temperature of B20-MnSi films

Zichao Li[1,2], Ye Yuan[1,3*], Viktor Begeza[1,4], Lars Rebohle[1], Manfred Helm[1,4], Kornelius Nielsch[2,4,5], Slawomir Prucnal[1], Shengqiang Zhou[1*]

[1]Helmholtz-Zentrum Dresden-Rossendorf, Institute of Ion Beam Physics and Materials Research, Bautzner Landstrasse 400, D-01328 Dresden, Germany

[2]Institute of Materials Science, Technische Universität Dresden, 01069 Dresden, Germany

[3]Songshan Lake Materials Laboratory, Dongguan, Guangdong 523808, People's Republic of China

[4]Institute of Applied Physics, Technische Universität Dresden, 01062 Dresden, Germany

[5]Institute for Metallic Materials, IFW-Dresden, Dresden, 01069, Germany

* Corresponding author: Ye Yuan, Shengqiang Zhou

E-mail address: yuanye@sslab.org.cn, s.zhou@hzdr.de

Abstract: B20-type MnSi is the prototype magnetic skyrmion material. Thin films of MnSi show a higher Curie temperature than their bulk counterpart. However, it is not yet clear what mechanism leads to the increase of the Curie temperature. In this work, we grow MnSi films on Si(100) and Si(111) substrates with a broad variation in their structures. By controlling the Mn thickness and annealing parameters, the pure MnSi phase of polycrystalline and textured nature as well as the mixed phase of MnSi and $MnSi_{1.7}$ are obtained. Surprisingly, all these MnSi films show an increased Curie temperature of up to around 43 K. The Curie temperature is likely independent of the structural parameters within our accessibility including the film thickness above a threshold, strain, cell volume and the mixture with $MnSi_{1.7}$. However, a pronounced phonon softening is observed for all samples, which can tentatively be attributed to slight Mn excess from stoichiometry, leading to the increased Curie temperature.

**Introduction**

Bulk manganese monosilicide (MnSi) is a weak itinerant helical magnet with B20 crystal structure [1]. At ambient pressure, bulk MnSi shows magnetic order with a Curie temperature ($T_C$) of ~ 29.5 K. It has been under investigation for a few decades regarding its intriguing physical properties, such as magnetic quantum phase transition [2, 3] and the formation of a non-Fermi liquid phase [4, 5]. With respect to practical applications, the most attractive property of MnSi is the formation of a magnetic skyrmion lattice, which is a topologically stable spin configuration and promising for spintronic application. A magnetic skyrmion lattice was experimentally observed in bulk MnSi by Mühlbauer *et al.* by using small angle neutron scattering [6]. This work has greatly motivated the development of MnSi thin films on Si substrates. Generally, MnSi thin films have been prepared by molecular beam epitaxy (MBE) [7-11], solid-state phase epitaxy [12-14] and magnetron sputtering [15]. Interestingly, independent of the preparation methods, epitaxial-like MnSi films grown on Si(111) show much enhanced $T_C$ to 35-45 K [11-13]. On Si(111) substrates, MnSi(111) is rotated by 30° with the orientation relationship of Si(111) ∥ MnSi(111) and Si[11$\bar{2}$] ∥ MnSi[1$\bar{1}$0], leading to a lattice mismatch of around -3.0% ($[a_{MnSi} \cos(30º) - a_{Si}]/a_{Si} = -3.0\%$). This induces an in-plane lattice expansion in the MnSi films. It has been shown experimentally that for bulk MnSi hydrostatic pressure decreases its $T_C$ [2] while negative chemical pressure can increase its $T_C$ [16, 17]. Therefore, the increased $T_C$ in MnSi films was presumably attributed to the tensile strain from the mismatch with Si substrate [10, 12]. However, the detailed analysis does not support this assumption, since thinner films show lower $T_C$ than thicker films [12, 18].

Karhu *et al.* have systematically checked the change of $T_C$ on MnSi films with different thicknesses and strain [12]. Indeed, it was found out that the thinner films show lower $T_C$ and all thicker (>10 nm) films exhibit a similar $T_C$ at around 43 K. A proportional correlation is observed between $T_C$ and the ratio between the out-of-plane and the in-plane strain. Li *et al.* also found that the $T_C$ of 50 nm thick MnSi film is almost identical to that of the 10 nm film [13]. López *et al.* found that a 30 nm-thick MnSi film does not develop any long-range magnetic order, while the 150 nm MnSi film has a $T_C$ at 34 K [15]. In general, it is known that with increasing thickness of the thin films, the strain originating from the interface should relax [19, 20]. In thicker MnSi films, the $T_C$ is expected to be lower than in thinner films. To understand the relationship between strain, atomic bonds and $T_C$ in MnSi films, Figueroa *et al.* have investigated thick MnSi films by polarization-dependent extended X-ray absorption fine structure and found that the Mn positions are unchanged. They concluded that for thick MnSi films the unit cell volume should be essentially the same as for bulk MnSi. They attributed the enhanced $T_C$ to the interface, whose particular unidentified characteristics strongly affect the magnetic properties of the entire MnSi film, even far from the interface. However, in other literature the very thin (below 5 nm) MnSi film shows lower $T_C$ [12, 18] or the absence of magnetic order [15]. Engelke et al. proposed that the film morphology may play a critical role [18]. Just recently, Sukhanov *et al.* reported an improved $T_C$ of bulk MnSi lamellae with μm dimensions embedded in $MnSi_x$ ( x~1.7) matrix [21]. The lattice mismatch between MnSi lamellae and the $MnSi_{1.7}$ matrix produces a tensile strain in MnSi. To understand the increased $T_C$, it has been assumed that the interface influences the μm thick lamellae. Therefore, the origin of the increased $T_C$ in MnSi films is still illusive.

Here, we report a systematic investigation on the Curie temperature of MnSi thin films with a large variation in their structural properties (see Table 1). These thin films were prepared by the solid-state reaction of metallic Mn layers with Si during ms-range flash lamp annealing. By controlling the Mn thicknesses and annealing parameters (energy density deposited to the sample surface by flash lamps), pure phase B20-MnSi and its mixture with $MnSi_{1.7}$ are prepared both on Si(100) and Si(111) substrates. All obtained thin films have a high Curie temperature around 43 K and the characteristic signature of magnetic skyrmions. We attempt to find a correlation between the Curie temperature and the structural properties, and therefore shed light on the understanding of the increased Curie temperature in thin films.

**Results**

Figure 1 (a) shows the XRD pattern of sample G with a 60 nm MnSi film grown on a Si(100) substrate. The (200) and (400) diffraction peaks of the Si substrate are at 33.05° and 69.2°, respectively. The MnSi (210) and (211) Bragg peaks are observed at 44.6° and 49.2°, respectively. According to the powder PDF card (n. 01-081-0484) [25], the MnSi (210) peak is the strongest. Taking into account the intensity ratio between different peaks, the MnSi film grown on Si(100) exhibits a polycrystalline nature. Furthermore, the $MnSi_{1.7}$ (104) peak appears at 26.03°. These two phases (MnSi and $MnSi_{1.7}$) often coexist [7, 26]. The XRD pattern of sample L with a MnSi film grown on a Si(111) substrate is shown in Fig. 1 (b). The diffraction peaks at around 28.4° and 58.9° are from the (111) and (222) of the Si substrate, respectively. The MnSi(111) and (222) peaks are observed at 34.2° and 72.4°, respectively. The MnSi(210) peak is also observed at 44.6°, but with much weaker intensity. Considering the intensity ratio between different peaks, the MnSi phase in this sample is highly (111) textured. Within the detection limit, there is no visible peak, which can be assigned to the $MnSi_{1.7}$. From our measurements for other samples with different thickness (not shown), we have found the following: (1) MnSi films on Si(100) are always polycrystalline with the co-existence of the MnSi$_{1.7}$ second phase; (2) MnSi films on Si(111) are highly (111) textured and we can obtain either pure MnSi phase or the mixture of two phases with different concentration ratios.

Table 1. The parameters of the samples and their Curie temperature ($T_C$).

| Sample ID | Substrate | Thickness of the regrown layer (nm) | Flash energy density (J/cm$^2$) | Anneal surface | Content of MnSi$_{1.7}$ (%) | $T_C$ (K) |
|---|---|---|---|---|---|---|
| A | Si(100) | 14 | 115 | Mn surface | 81 | 37±2 |
| B | Si(100) | 20 | 115 | Mn surface | 60 | 39±2 |
| C | Si(100) | 30 | 140 | Si surface | 66 | 42±2 |
| D | Si(100) | 40 | 115 | Mn surface | 65 | 45±1 |
| E | Si(100) | 60 | 135 | Si surface | 85 | 44±1 |
| F | Si(100) | 60 | 140 | Si surface | 77 | 43±2 |
| G | Si(100) | 60 | 140 | Mn surface | 57 | 44±2 |
| H | Si(111) | 30 | 140 | Si surface | 85 | 39±2 |
| I | Si(111) | 40 | 110 | Mn surface | 67 | 44±2 |
| J | Si(111) | 40 | 115 | Mn surface | 53 | 46±1 |
| K | Si(111) | 60 | 140 | Mn surface | 73 | 42±1 |
| L | Si(111) | 60 | 140 | Si surface | 0 | 45±2 |

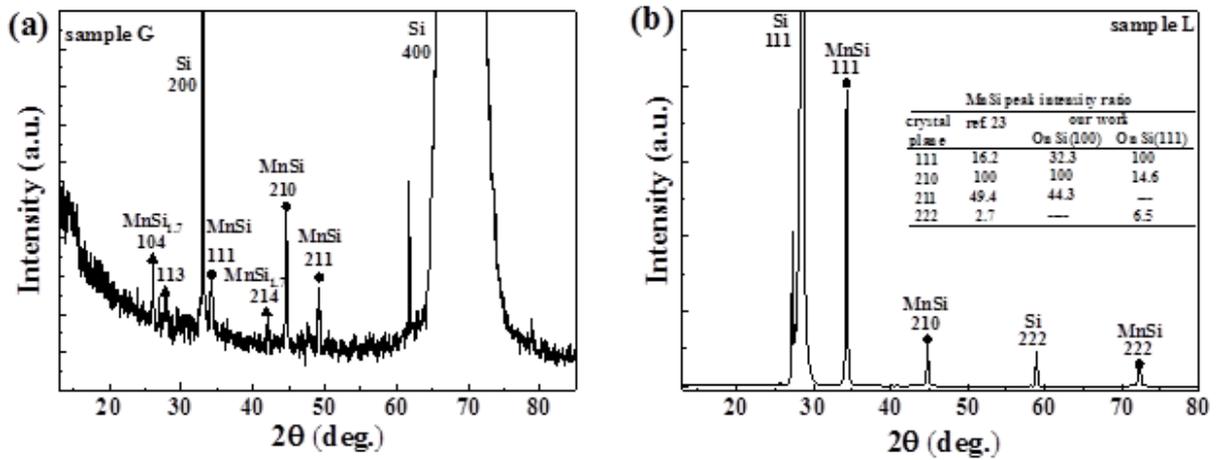

Figure 1. (a) XRD pattern of a 60 nm MnSi film on Si(100) by FLA. B20-MnSi and MnSi$_{1.7}$ phases coexist. (b) XRD pattern of a 60 nm MnSi film on Si(111) by FLA, and in this sample B20-type MnSi is the single phase. The insert table in (b) shows the relative intensity ratio of different diffraction planes. Ref. 23 shows the (210) and (211) planes should be the two strongest peaks. Sample G shows consistent with Ref. 23, indicating a polycrystalline structure. The (111) plane of sample L is the strongest peak, meaning the (111)-textured of this sample.

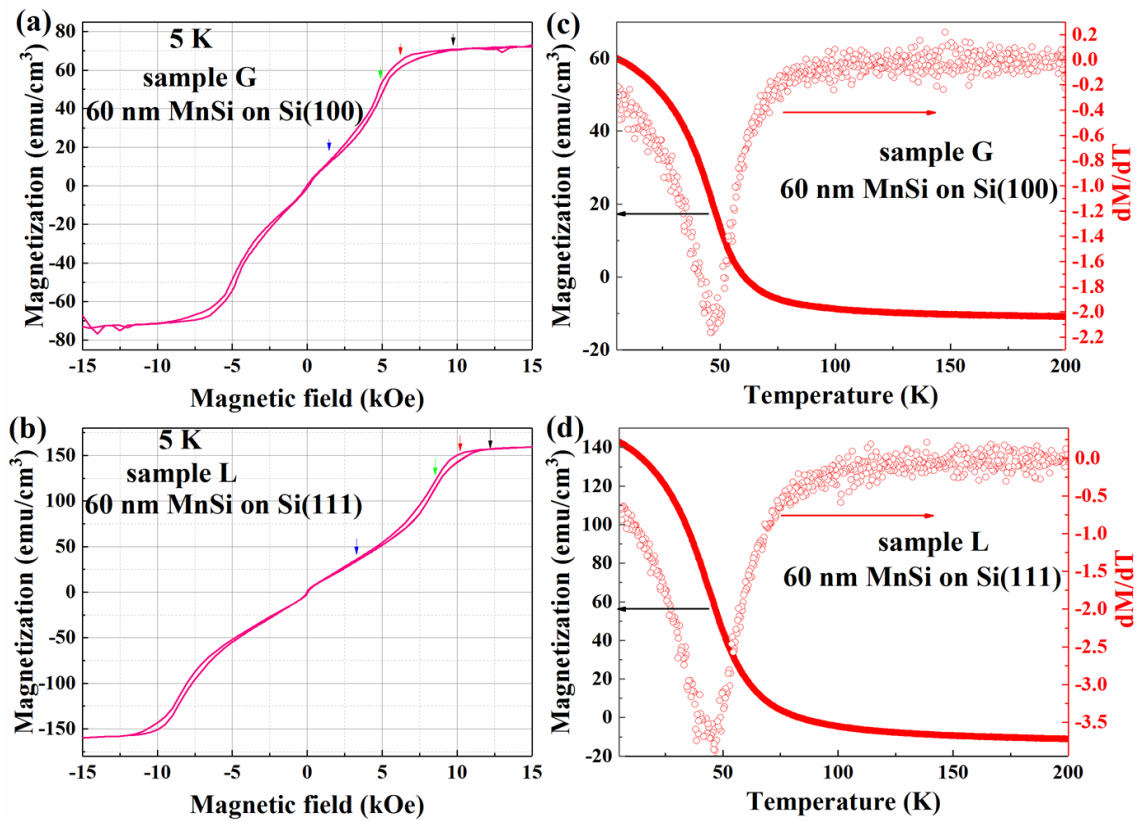

Figure 2. In-plane MH curves recorded at 5 K for samples G and L with a 60 nm MnSi films on (a) Si(100) substrates and (b) on Si(111) substrates, respectively. The easy axis and multi-hysteresis are stabilized in-plane. The arrows obtained from the peaks of the differential MH curves indicate the fields where the magnetic phase

transition occurs. Temperature-dependent in-plane saturation magnetization (solid symbol) and the calculated *dM/dT* (open symbol) for samples G (c) and L (d). The valley of *dM/dT* indicates the Curie temperature.

As shown exemplarily in Fig. 2 (a) and (b) for samples G and L with 60 nm MnSi films grown on Si(100) and (111) substrates, the in-plane magnetic loops at 5 K show a multi-hysteresis feature that occurs only when the magnetic states change, which is similar to the results reported for other MnSi films [27] or bulk [21]. It is consistent with the magnetic helicoidal states at low temperature as suggested by Wilson *et al* [28]. Later, magnetic skrymions were found at higher temperatures by the same research group [29]. The saturation magnetization for sample G is around 75 emu/cm$^3$, which is lower than the value for bulk B20 MnSi due to the co-existence of MnSi$_{1.7}$ parasitic phase. The saturation field for this sample is around 9 kOe. However, the single-phase sample L in Fig. 2 (b) shows a higher saturation magnetization of 165 emu/cm$^3$, which is close to that for bulk MnSi. This sample also has a higher saturation field of 13 kOe. The mixture with MnSi$_{1.7}$ does not affect this multi-hysteresis behavior. However, there is a clear difference in the critical fields (obtained from the peaks of the differential MH curves) labelled by arrows in the figure. This hints towards different magnetic anisotropy in the films on Si(100) and (111) substrates. The fully skyrmions are stabilized in the range between green and red arrows.

The Curie temperature of magnetic materials can be measured by different methods, such as temperature dependent magnetization, resistance or heat capacity [30-32]. From the magnetization, it can be determined by the temperature-dependent remanence [33] and by calculating *dM/dT* for the temperature-dependent magnetization [12]. $T_C$ can also be determined by measuring the temperature-dependent resistance. In such a case the first derivative *dR/dT* shows a peak around the critical temperature [34]. At the same time, the temperature-dependent magnetoresistance also shows a peak around $T_C$ [35]. Fig. 2 (c) and (d) show the in-plane magnetization of samples G and L as a function of temperature under a magnetic field of 15 kOe, which is above the saturation field. The calculated *dM/dT* curves are shown as open symbols. The minimum of the curves is found at the same temperature for films on Si(100) and (111) substrates, which is defined as the Curie temperature. We also estimated the $T_C$ from electrical measurements (not shown). Both methods, i.e. temperature-dependent magnetization and resistance result in a similar value of $T_C$ at around 43 K.

The $T_C$ values are plotted in Fig. 3 (a) and (b) as a function of MnSi film thickness. The $T_C$ slightly increases to 43 K with increasing thickness of the MnSi films on Si(100) or (111) substrates. When the thickness is above 30 nm, $T_C$ saturates at around 43 K. Compared with bulk MnSi, it is increased by 45% and remains stable at 43 K. For thinner samples, $T_C$ is a little bit lower, consistent with Ref. 12. Although these samples have different crystal orientations or textures on Si(100) or (111) substrates, their $T_C$ is almost the same, indicating that the $T_C$ of MnSi films is not related to the crystal orientation or texture.

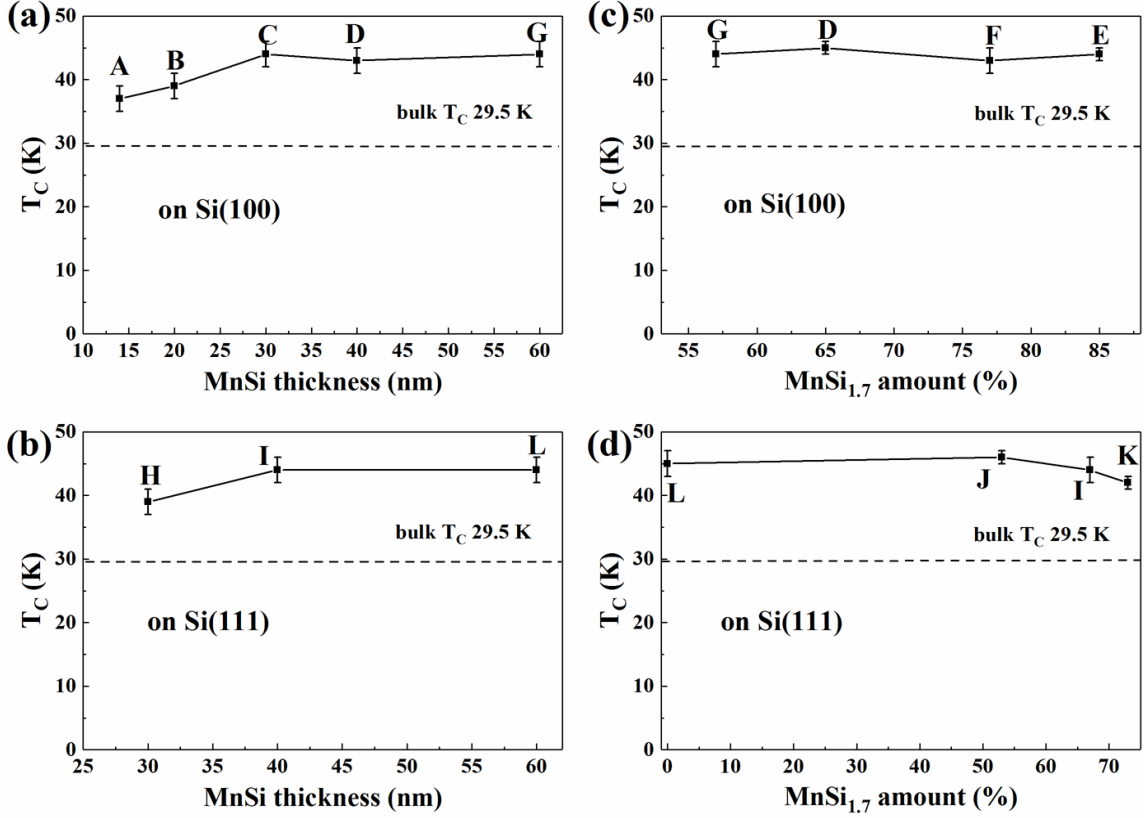

Figure 3. Curie temperature as a function of the thickness of MnSi films on (a) Si(100) substrates and Si(111) substrates (b). Curie temperature as a function of the content of $MnSi_{1.7}$ phase on (c) Si(100) substrates and Si(111) substrates (d). All these samples have approximately the same Curie temperature of 43 K, which is larger than that of bulk MnSi. The sample ID is indicated in the figures around the corresponding data point.

In order to check if the presence of $MnSi_{1.7}$ has influence on the $T_C$ of MnSi films [21], we tried to find a correlation between $T_C$ and the content of $MnSi_{1.7}$. The content of $MnSi_{1.7}$ in percent is estimated from the saturation magnetization. Since $MnSi_{1.7}$ is a weak itinerant magnet with negligible magnetization of 0.012 $\mu_B$/Mn [24], we compare the saturation magnetization of our films with that of bulk MnSi. Then the amount of MnSi is determined and we assume that the rest of Mn form $MnSi_{1.7}$. As shown in Fig. 3 (c, d), the $T_C$ always remains around 43 K with increasing amount of the $MnSi_{1.7}$ phase on Si(100) or (111) substrates. Here, the film thickness is fixed at around 40 and 60 nm and the amount of $MnSi_{1.7}$ is varied since we changed the FLA energy slightly [22]. Therefore, $T_C$ of MnSi films is not determined by its crystalline orientation, texture or the mixture with MnSi$_{1.7}$. In the next section, we investigate the dependence on strain.

Thin films on substrates are often in a "stressed" state. The strain of thin films can be calculated by the variation of lattice constants. The change of the lattice constant '*a*' and '*c*' is generally regarded as in-plane and out-of-plane strain, respectively [36], in the case of epitaxy. The change of the lattice constant can be obtained by XRD measurements. The lattice spacing *d* can be calculated from equation (1) of Bragg's law. $\theta$ is the Bragg peak of the specific crystal plane, which can be expressed as the Miller index (hkl). $\lambda$ is the wavelength of the incident X-ray. *n* is the diffraction order and is a positive integer.

$$2d\sin\theta = n\lambda \tag{1}$$

The lattice constant '*a*' for a cubic crystal can be obtained as follows:

$$\frac{1}{d^2} = \frac{h^2+k^2+l^2}{a^2} \tag{2}$$

For the thin film grown on Si(100) substrates, MnSi is fully polycrystalline because the XRD intensity ratios for different diffraction peaks are the same as for the powder sample. Thus, we can calculate the change of average lattice constants by equation (2) as for polycrystalline samples [37].

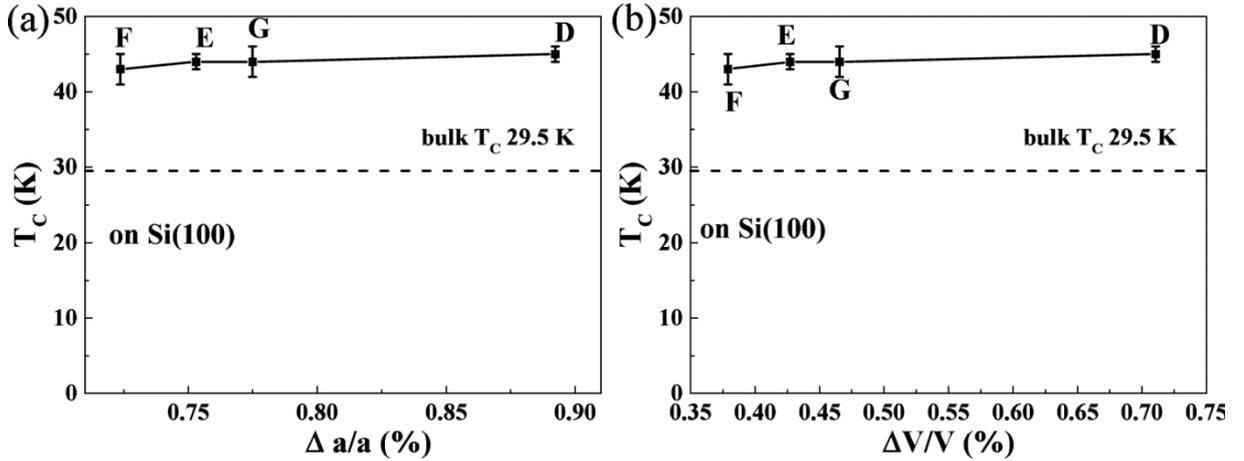

Figure 4. (a) Curie temperature as a function of the change of lattice constant '*a*' for MnSi films on Si(100) substrates. (b) The Curie temperature vs. the change of the cell volume of the same set of MnSi films. The Curie temperature basically stays around 43 K regardless the change of lattice constant or cell volume. The sample ID is indicated in the figures around the corresponding data point.

Figure 4(a) shows the Curie temperature dependent on the variation of lattice constants '*a*' for the MnSi films on Si(100) substrates. These MnSi films have a thickness of 40 or 60 nm.

The lattice constants vary due to the different flash lamp annealing energy. As shown in Fig. 4 (a), the Curie temperatures of all MnSi films stay around 43 K. In the case of growing on Si(100), the lattice constant *a* is larger than the bulk values. However, for all cases the $T_C$ of MnSi films is around 43 K and is not determined by the change of the lattice constant '*a*', which is somewhat equivalent to the strain. The Curie temperature dependence on the volume variation of MnSi films is shown in Fig. 4 (b). Obviously, independent of the lattice cell expansion values, MnSi films have a $T_C$ around 43 K. MnSi films grown on Si(111) substrates are partially (111) textured as shown in Fig. 1(b). If we assume the (111) and (222) diffraction peaks stem solely from those (111) textured crystallites, we can estimate their out-of-plane strain to lie between -0.1% and 0.3% for samples I, J, K, L. However, we are not able to measure reliable XRD results to estimate the in-plane strain. Nevertheless, the Curie temperature for samples I, J, K, and L shows a similar value around 43 K and has no dependence on their strain status.

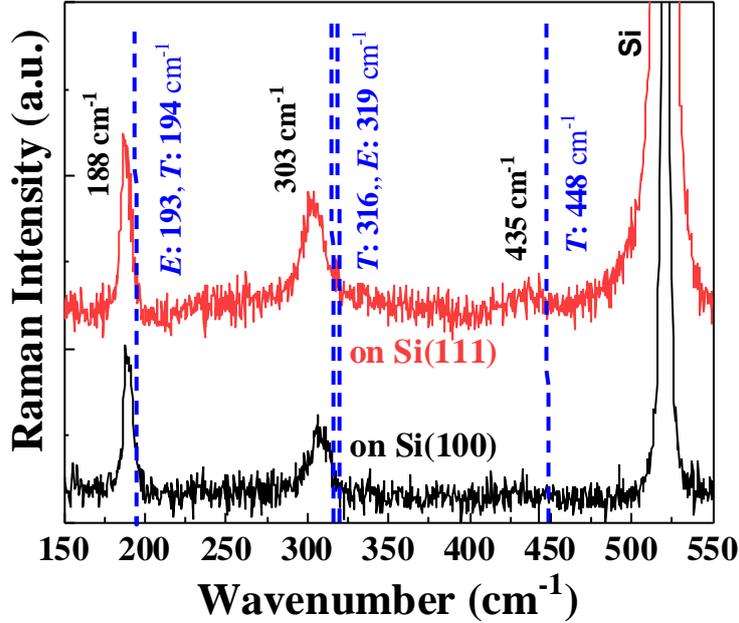

Figure 5. Representative room-temperature Raman spectra for different samples on Si(100) (black) and (111) (red) substrates. One spectrum is vertically shifted for better visibility. For both samples, the Raman modes shift to lower wavenumbers. The known Raman frequencies (in cm$^{-1}$) of (*E*, *T*) modes for bulk MnSi are indicated by the dash lines (blue) in the figure for comparison.

To further scrutinize the origin of the increase $T_C$ in our MnSi films, we employed Raman scattering, which is a fast and non-destructive method to measure the phonon frequencies in materials with possible spatial resolution below micrometers [38, 39]. For bulk B20-MnSi, nine optical phonon modes (2*A*+2*E*+5*T*) from *A*, *E* and *T* symmetries are Raman active depending on the measurement geometry [40, 41]. Among them, the Raman peaks from the *A* symmetry are relatively weak [42]. The representative Raman spectra of samples on Si(100) and (111) are shown in Fig. 5. Two prominent Raman peaks appear at around 188 cm$^{-1}$ and 303 cm$^{-1}$, respectively. For bulk MnSi crystals, the active Raman vibrations in these spectral ranges can be $E$ and $T$ modes with frequencies of 193, 194 cm$^{-1}$ and 319, 316 cm$^{-1}$ at room temperature, respectively [40, 41]. Since our samples are quasi-polycrystalline, we cannot unambiguously assign them to $E$ or $T$ modes. However, the Raman modes from our films are red shifted regardless their mode assignments, including the weak Raman mode at around 435 cm$^{-1}$. The red shifts amount to about 5-6 cm$^{-1}$ or 13-16 cm$^{-1}$ for the two prominent peaks, respectively. In Fig. 6, we map all investigated samples by plotting T$_C$ as a function of Raman shifts. Note that, for samples A, B and C, the MnSi layers are too thin and the second Raman peak around 303 cm$^{-1}$ is not resolvable. A reference sample prepared by molecular beam epitaxy from Ref. [42] is also included for comparison. The Curie temperature of MnSi films does not reveal a clear dependence on the Raman shift.

In general, the red-shift in Raman modes can be due to tensile strain or a variation of composition in compounds. And with increasing carrier concentration (the decrease of resistivity), the electron-phonon coupling and electronic screening will also cause the phonon peak shift to a lower wavenumber. In our case, the films have slightly higher resistivity than bulk MnSi: ~ 28×10$^{-6}$ Ωcm for our films [22] vs. 3-4×10$^{-6}$ Ωcm for bulk MnSi [2] at 5 K. This result is in agreement with Ref. [43], in which the authors also measured a much smaller carrier concentration in MnSi films than in bulk crystal. Therefore, the red-shift induced by the electron-phonon coupling and electronic screening due to a high carrier concentration in MnSi films is not expected.

Now we discuss the influence from the strain or variation of composition. The solid solutions GeSi, GeSn are typical examples in which a concentration increase of heavy elements leads to a large red-shift in Raman spectra [44]. For GeSi and GeSn, it is possible to quantitatively correlate the Raman shift to strain or composition by comprehensive XRD and Raman measurements since many material parameters are known [44]. However, for MnSi there are very limited data available in literature, preventing a quantitative analysis. In analogy to the case of GeSn with a prominent Raman mode around 300 cm$^{-1}$, we try to make a qualitative estimation. If the red-shift in MnSi films were solely due to strain, it would mean that our films are under > 5% in-plane tensile strain yielding a red shift of around 10 cm$^{-1}$, i.e. from 316 cm$^{-1}$ to around 303-307 cm$^{-1}$. This is unrealistically high and much higher than the estimated value from XRD for the polycrystalline films (shown in Fig. 4). A more plausible interpretation is a slight Mn excess, i.e. the stoichiometry is $Mn_{0.5+x}Si_{0.5-x}$ instead $Mn_{0.5}Si_{0.5}$, since Mn is heavier than Si. A Mn excess of a few percent could result in 5-10 cm$^{-1}$ red-shift. However, we cannot quantify the off-stoichiometry due to the limited data available in literature. Nevertheless, there is no clear correlation between $T_C$ and the Raman shift as shown in Fig. 6.

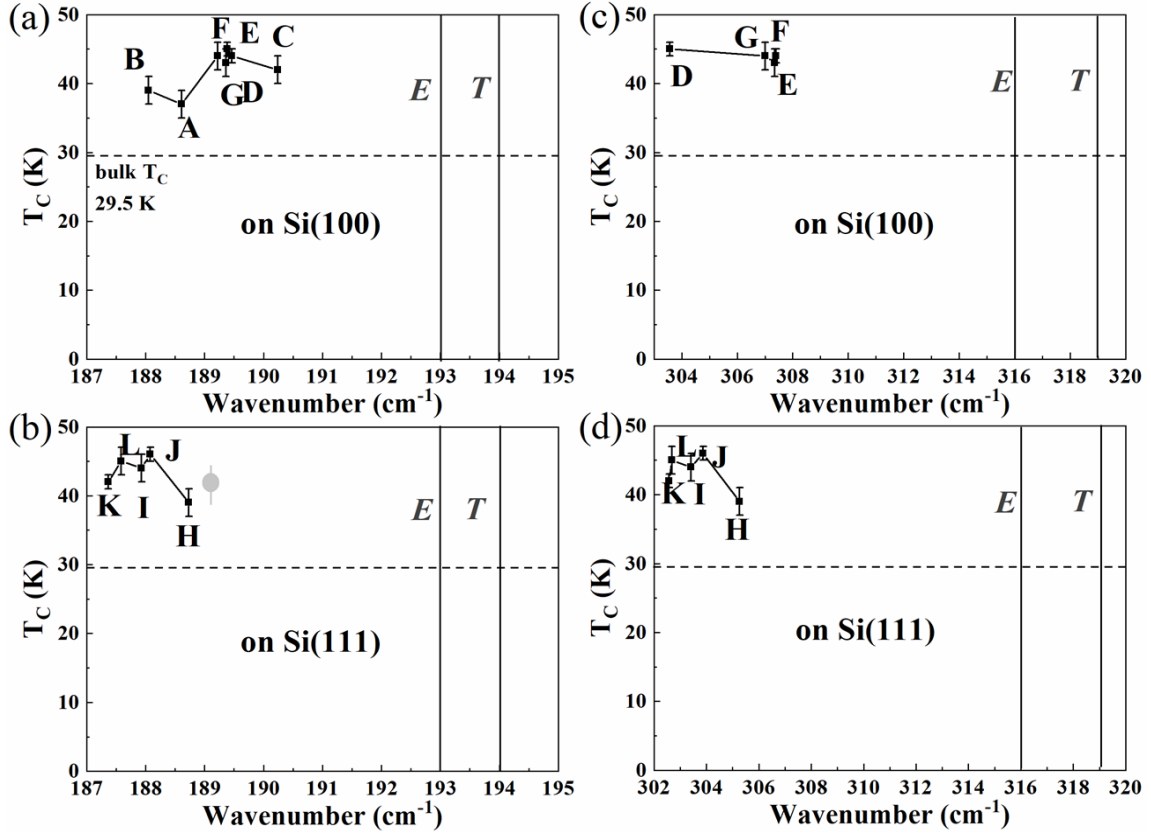

Figure 6. Curie temperature as a function of the Raman shifts of MnSi films. (a) and (b) for the Raman mode around 188 cm$^{-1}$. (c) and (d) for the Raman mode around 303 cm$^{-1}$. All samples have approximately the same Curie temperature and show no dependence on the Raman shift. The sample ID is indicated in the figures around the corresponding data point. In panel (c) the data point indicated by a grey solid circle is from Ref. [41]. The dashed horizontal lines labels the Curie temperature of 29.5 K for bulk MnSi. The vertical solids labels the Raman modes for bulk MnSi.

**Discussion**

From the experimental results shown above, the origin of the increased $T_C$ in MnSi films remains still elusive. We found that, being independent of the crystallinity, the substrate orientation, the lattice strain and the cell volume, $T_C$ is always around 43 K when the thickness is above a threshold value. By detailed investigations using EXAFS, Figueroa *et al.* suggested that the interface may be the reason for the improved Curie temperature. They also suspected that the shifted Si positions within the unit cell could have an indirect effect on the magnetic ordering of the Mn atoms, e.g., via the crystal field. However, they have not detected skyrmions in their films. In our thick films, both magnetic skyrmions and an enhanced Curie temperature are detected [22]. Therefore, the understanding of the magnetic properties of MnSi films is far from satisfactory. Instead, subtle variations in the microstructure may play an important role. Both theoretical works by Yabuuchi *et al.* [45] and Men'shov *et al.* [46] point to off-stoichiometry induced spin fluctuations above the intrinsic Curie temperature in MnSi alloys. In experiments, Rylkov *et al.* have found that in Si-Mn films ($MnSi_{1.7}$ and MnSi) with a slight excess of Mn (around 2-5% from their stoichiometry) the Curie temperature increases [47, 48]. However, the skyrmion formation was not discussed in those publications. Very recently, Balasubramanian *et al.* reported ferromagnetic order in B20-CoSi [49]. Perfectly stoichiometric CoSi does not exhibit any kind of magnetic order. $B20-Co_{1+x}Si_{1-x}$ with excess Co was obtained by nonequilibrium processing. They found that the alloys are magnetically ordered above a critical excess-Co content (x=0.028). Their density functional theory calculation shows that the onset of the zero-temperature magnetism has the character of a magnetic quantum-phase transition. This is in line with our Raman observation for MnSi films. Very likely there is slightly more Mn contained in the MnSi films, inducing the red-shift in Raman and the increased Curie temperature. Nevertheless, the inevitable subtle deviation in stoichiometry and point defects in MnSi films seem to play a non-negeligible role in their magnetic properties. Indeed, it is challenging to quantitatively characterize the subtle amount of off-stoichiometry and point defects. To tackle this problem, well-controlled growth of MnSi films and sensitive characterization are pre-requisites. For the latter, positron annihilation spectroscopy may provide more information about the defects [50].

**Conclusion**

In summary, in this paper we try to understand the puzzling enhancement of Curie temperature widely reported in MnSi films. We have prepared MnSi films with a large variation regarding their thickness, crystallinity, strain and phase separation by a fast solid-state reaction though millisecond flash lamp annealing. Particularly, polycrystalline MnSi films on Si(100) and textured MnSi films on Si(111), both with different mixture ratio with $MnSi_{1.7}$ have been grown and systematically characterized. Surprisingly, all obtained MnSi films exhibit a high Curie temperature at around 43 K when the thickness is above a threshold value. However, we find no correlation between the increased Curie temperate and the film thickness (above 30 nm), strain, lattice volume or the mixture with $MnSi_{1.7}$. Instead, we find indications of Mn excess in all films by Raman spectroscopy, which may be responsible for the increased Curie temperature. Although our work has not provided a conclusive picture for this question, it is rather calling for a revisiting, especially concerning the effects of interface, stoichiometry and point defects. Further studies are essential to understand the B20 transition-metal silicide/germanides films and therefore to utilize them for spintronic applications.

**Methods**

To prepare MnSi films, 7-30 nm thick Mn films were firstly deposited on Si(100) and (111) wafers by DC magnetron sputtering. Afterwards, flash lamp annealing (FLA) was employed to realize a fast solid-state reaction between Mn and Si at different annealing parameters. The largest thickness of the MnSi is about 60 nm (see Table 1). During the FLA process, these samples were heated up by 12 Xe-lamps in a continuous $N_2$ flow. Samples were annealed either from the front side or from the backside with a peak temperature above 1300 K. With a 20 ms pulse duration, the heating and cooling rates are estimated to be around 80000 and 160 $Ks^{-1}$, respectively. Such high heating /cooling rate will allow the control over the parasitic growth of $MnSi_{1.7}$ in B20-type MnSi. By changing the flash lamp energy (and therefore the peak temperature), we can selectively prepare the pure phases of MnSi and $MnSi_{1.7}$ or their mixture. The details about the preparation have been reported in Ref. 22. All samples used in this manuscript are summarized in Table 1. The saturation magnetization of B20-MnSi and $MnSi_{1.7}$ is 0.4 $\mu_B$/Mn and 0.012 $\mu_B$/Mn, respectively [23, 24]. We calculated the amount of MnSi according to the saturation magnetization by neglecting the magnetization of $MnSi_{1.7}$.

X-ray diffraction (XRD) was employed to analyse the microstructure of the obtained films. XRD was performed at room temperature on a Bruker D8 Advance diffractometer with a Cu-target source. The measurements were done in Bragg-Brentano-geometry with a graphite secondary monochromator and a scintillator. Micro-Raman spectroscopy was employed to measure the possible strain and stoichiometry deviation. The experiments were performed using a Horiba micro-Raman system with the excitation wavelength of 532 nm and the signal was recorded with a liquid-nitrogen-cooled silicon CCD camera. The magnetic properties of the films were measured by a superconducting quantum interference device equipped with a vibrating sample magnetometer (SQUID-VSM) with the field parallel (in-plane) to the films. For measuring the temperature-dependent magnetization (MT), the samples were cooled down to 5 K under a zero field, then a 15 kOe magnetic field was applied and the magnetization data was collected during the warming up process. The transport properties of MnSi films were investigated by a Lake Shore Hall measurement system. The temperature-dependent resistance was measured from 5 to 300 K using the van der Pauw geometry.

**Data availability**

All data generated or analyzed during this study are included in this published article, and the datasets used and analyzed during the current study are available from the corresponding author on reasonable request.

**Acknowledgment**

Z. Li thanks the assistance by Andrea Scholz and Tianbing He for XRD measurements. The authors are grateful to Ruben Hühne and Jörg Grenzer for the fruitful discussion about XRD measurements and analysis. Z. L. acknowledges the financial support by China Scholarship Council (File No. 201707040077). The work is also partially supported by German Research Foundation (DFG, ZH 225/6-1) and the Basic and Application Basic Research Foundation of Guangdong Province (Grant No. 2020A1515110891).

**Author contribution statement**

Y. Y. and Z. L. initiated and designed experiments. S. Z. and Z. L. wrote the manuscript from integrating input data and analyses provided from all the authors. V. B., L. R., S. P. and Z. L. synthesized samples and performed Raman measurements; Z. L. performed magnetic and transport measurements. S. Z., K. N. and M. H. supervised the research.

**Competing interests**

The authors declare no competing interests.

Figure captions

Figure 1. (a) XRD pattern of a 60 nm MnSi film on Si(100) by FLA. B20-MnSi and MnSi$_{1.7}$ phases coexist. (b) XRD pattern of a 60 nm MnSi film on Si(111) by FLA, and in this sample B20-type MnSi is the single phase. The insert table in (b) shows the relative intensity ratio of different diffraction planes. Ref. 23 shows the (210) and (211) planes should be the two strongest peaks. Sample G shows consistent with Ref. 23, indicating a polycrystalline structure. The (111) plane of sample L is the strongest peak, meaning the (111)-textured of this sample.

Figure 2. In-plane MH curves recorded at 5 K for samples G and L with a 60 nm MnSi films on (a) Si(100) substrates and (b) on Si(111) substrates, respectively. The easy axis and multi-hysteresis are stabilized in-plane. The arrows obtained from the peaks of the differential MH curves indicate the fields where the magnetic phase transition occurs. Temperature-dependent in-plane saturation magnetization (solid symbol) and the calculated *dM/dT* (open symbol) for samples G (c) and L (d). The valley of *dM/dT* indicates the Curie temperature.

Figure 3. Curie temperature as a function of the thickness of MnSi films on (a) Si(100) substrates and Si(111) substrates (b). Curie temperature as a function of the content of MnSi$_{1.7}$ phase on (c) Si(100) substrates and Si(111) substrates (d). All these samples have approximately the same Curie temperature of 43 K, which is larger than that of bulk MnSi. The sample ID is indicated in the figures around the corresponding data point.

Figure 4. (a) Curie temperature as a function of the change of lattice constant '*a*' for MnSi films on Si(100) substrates. (b) The Curie temperature vs. the change of the cell volume of the same set of MnSi films. The Curie temperature basically stays around 43 K regardless the change of lattice constant or cell volume. The sample ID is indicated in the figures around the corresponding data point.

Figure 5. Representative room-temperature Raman spectra for different samples on Si(100) (black) and (111) (red) substrates. One spectrum is vertically shifted for better visibility. For both samples, the Raman modes shift to lower wavenumbers. The known Raman frequencies (in cm$^{-1}$) of (*E*, *T*) modes for bulk MnSi are indicated by the dash lines (blue) in the figure for comparison.

Figure 6. Curie temperature as a function of the Raman shifts of MnSi films. (a) and (b) for the Raman mode around 188 cm$^{-1}$. (c) and (d) for the Raman mode around 303 cm$^{-1}$. All samples have approximately the same Curie temperature and show no dependence on the Raman shift. The sample ID is indicated in the figures around the corresponding data point. In panel (c) the data point indicated by a grey solid circle is from Ref. [41]. The dashed horizontal lines labels the Curie temperature of 29.5 K for bulk MnSi. The vertical solids labels the Raman modes for bulk MnSi.

Table 1. The parameters of the samples and their Curie temperature ($T_C$).

| Sample ID | Substrate | Thickness of the regrown layer (nm) | Flash energy density (J/cm$^2$) | Anneal surface | Content of MnSi$_{1.7}$ (%) | $T_C$ (K) |
|---|---|---|---|---|---|---|
| A | Si(100) | 14 | 115 | Mn surface | 81 | 37±2 |
| B | Si(100) | 20 | 115 | Mn surface | 60 | 39±2 |
| C | Si(100) | 30 | 140 | Si surface | 66 | 42±2 |
| D | Si(100) | 40 | 115 | Mn surface | 65 | 45±1 |
| E | Si(100) | 60 | 135 | Si surface | 85 | 44±1 |
| F | Si(100) | 60 | 140 | Si surface | 77 | 43±2 |
| G | Si(100) | 60 | 140 | Mn surface | 57 | 44±2 |
| H | Si(111) | 30 | 140 | Si surface | 85 | 39±2 |
| I | Si(111) | 40 | 110 | Mn surface | 67 | 44±2 |
| J | Si(111) | 40 | 115 | Mn surface | 53 | 46±1 |
| K | Si(111) | 60 | 140 | Mn surface | 73 | 42±1 |
| L | Si(111) | 60 | 140 | Si surface | 0 | 45±2 |